\documentclass[12pt,a4paper]{article}
\usepackage{booktabs}
\usepackage{graphicx}
\usepackage{epsfig}
\usepackage{times}
\usepackage[english]{babel}
\makeatletter
\DeclareRobustCommand{\textsupsub}[2]{{%
  \m@th\ensuremath{%
    ^{\mbox{\fontsize\sf@size\z@#1}}%
    _{\mbox{\fontsize\sf@size\z@#2}}%
  }%
}}
\makeatother
\title{\bf Investigating Extra-solar Planetary System Qatar-1 through Transit Observations}
\author{Parijat Thakur$^1$\thanks{E-mail: parijat@associates.iucaa.in or parijatthakur@yahoo.com}, Vineet Kumar Mannaday$^1$, Ing-Guey Jiang$^2$, D. K. Sahu$^3$\\  and Swadesh Chand$^1$\\
\vspace{1cm}\\
\normalsize $^1$ Department of Pure \&\ Applied Physics, Guru Ghasidas Central University,\\ \normalsize Bilaspur (C.G.) – 495 009, India\\ 
\normalsize $^2$ Department of Physics and Institute of Astronomy, National Tsing-Hua University,\\ \normalsize Hsinchu, Taiwan \\
\normalsize $^3$ Indian Institute of Astrophysics, Bangalore – 560 043, India}
\date{\mbox{}}
\begin{document}
\maketitle
\pagestyle{empty}
%
%
\def\bull{\vrule height .9ex width .8ex depth -.1ex}
\makeatletter
\def\ps@plain{\let\@mkboth\gobbletwo
\def\@oddhead{}\def\@oddfoot{\hfil\scriptsize\bull\quad
"First Belgo-Indian Network for Astronomy \& astrophysics (BINA) workshop'', held in Nainital (India), 15-18 November 2016 \quad\bull}%
\def\@evenhead{}\let\@evenfoot\@oddfoot}
\makeatother
%
%
\def\beginrefer{\section*{References}%
\begin{quotation}\mbox{}\par}
\def\refer#1\par{{\setlength{\parindent}{-\leftmargin}\indent#1\par}}
\def\endrefer{\end{quotation}}
%
%
{\noindent\small{\bf Abstract:} 
We report the results of the transit timing variation (TTV) analysis of the extra-solar planet Qatar-1b using thirty eight light curves. Our analysis combines thirty five previously available transit light curves with three new transits observed by us between June 2016 and September 2016 using the 2-m Himalayan Chandra Telescope (HCT) at the Indian Astronomical Observatory (Hanle, India). From these transit data, the physical and orbital parameters of the Qatar-1 system are determined. In addition to this, the ephemeris for the orbital period and mid-transit time are refined to investigate the possible TTV. We find that the null-TTV model provides the better fit to the (O-C) data. This indicates that there is no evidence for TTVs to confirm the presence of additional planets in the Qatar-1 system. The use of the 3.6-m Devasthal Optical Telescope (DOT) operated by the Aryabhatta Research Institute of Observational Sciences (ARIES, Nainital, India) could improve the photometric precision to examine the signature of TTVs  in this system with a greater accuracy than in the present work.}
%
%
\section{Introduction}
The development of the research in extra-solar planets has been successful for nearly two decades. The Doppler-shift method allowed us to detect many extra-solar planets. In addition to this, the method of planetary transits also produces fruitful results. So far, already more than 2653 extra-solar planetary systems have been found to transit their parent stars. In order to study the perturbation from small unknown planets and to constrain the overall orbital configuration in planetary systems, TTVs have been seriously investigated in recent years (Agol et al. 2005; Holman et al. 2005, 2010; Maciejewski et al. 2015). Among the detected planetary systems, Qatar-1 attracts a lot of attention due to its strong transit signal and short orbital period. For the Qatar-1 system, Von Essen et al. (2013) have found evidence for possible TTVs, whereas some other workers in this field (e.g. Maciejewski et al. 2015; Collins et al. 2017) have not claimed the detection of TTVs. These results imply that further photometric follow-up of transits for the extra-solar planetary system Qatar-1 is necessary to confirm the presence or absence of TTVs. Here, in addition to three new transit observations of this system with the 2-m HCT telescope, we try to cover much more epochs by including many transit data from both the literature and  the Exoplanet Transit Database (ETD) \footnote {http://var.astro.cz/ETD} to improve the estimates of the physical and orbital parameters of this system as well as to refine the ephemeris for the orbital period and mid-transit time required for future transit observations.

The remainder of this paper is organized as follows. In the section 2, we describe the observations and data reduction. Section 3 presents the methods for analysis of transit light curves. Section 4 and 5 are devoted to the estimation of new ephemeris and TTV analysis, respectively. Finally, the concluding remarks are given in Section 6.
\section{Observation and Data Reduction}
During June-September 2016, we carried out the three photometric follow-up observations of the Qatar-1 system in the R-band by gathering 60 sec exposures with the 2-m HCT telescope. In addition to these new data, we also considered 8 light curves from ETD, 5 from Covino et al. (2013), 7 from Von Essen et at. (2013), and 15 from Maciejewski et al. (2015). In total, 38 light curves are analyzed in this study. The HCT CCD images of the Qatar-1 system are calibrated using the standard IRAF \footnote {IRAF (Image Reduction and Analysis Facility) is distributed by the National Optical Astronomy
Observatories, which are operated by the Association of Universities for Research in Astronomy, Inc., under cooperative agreement with the
National Science Foundation. For more details, http://iraf.noao.edu/} procedures such as trimming, bias subtractions, and flat field division. After pre-processing, aperture photometry has been  performed on the Qatar-1 system and the nearby comparison stars using the {\it {'phot'} task} within IRAF. Using the flux of Qatar-1 and the comparison stars, we carry out the differential photometry to construct the light curve of each observed transit.
\section{Analysis of Light Curves}
The Transit Analysis Package (TAP), as described in Gazak et al. (2012), has been used for our light curves analysis. The TAP employs Markov Chain Monte Carlo (MCMC) technique (see Gazak et al. 2012) and the model of Mandel \&\ Agol (2002) to fit the light curves. The results derived from our three HCT data sets are shown in Table 1, where 2\,457\,000 is already subtracted from the mid-transit time T\textsupsub{}{m}. The observed light curves with the modeled light curves are shown in Fig.1.
The orbital inclination i, the planet-to-star radius ratio R\textsupsub{}{p}/R\textsupsub{}{$\ast$} and the scaled semi-major axis a/R\textsupsub{}{$\ast$} are plotted  as a function of epoch in Fig. 2. It shows that there are no significant variations  in these parameters with the epochs as most of the points are confined within the $2\sigma$ levels.
\section{New Ephemeris}
We determined refined ephemeris for orbital and mid-transit time using the linear function, T\textsupsub{c}{m} = EP+T\textsupsub{}{0}, where T\textsupsub{c}{m}, E, P, and T\textsupsub{}{0} are  the calculated mid-transit time, epoch, orbital period, and mid-transit time at E = 0, respectively. The first Qatar-1 transit in Covino et al. (2013) is defined as epoch E = 0 and the other transits epochs are calculated accordingly. We have done the $\chi$\textsupsub{2}{} fitting on all the 38 data sets of (E, T\textsupsub{}{m}) and calculated the best fit values of P and T\textsupsub{}{0} with $\chi$\textsupsub{2}{red}= 1.12 as \\ P=1.42002368 $\pm$ 0.00000015 d and {T\textsupsub{}{0}} =2455711.53470 $\pm$ 0.00011 BJD{\textsupsub{}{TDB}}\\ \\
\begin{figure}[h]
\begin{center}
\includegraphics[width=8cm]{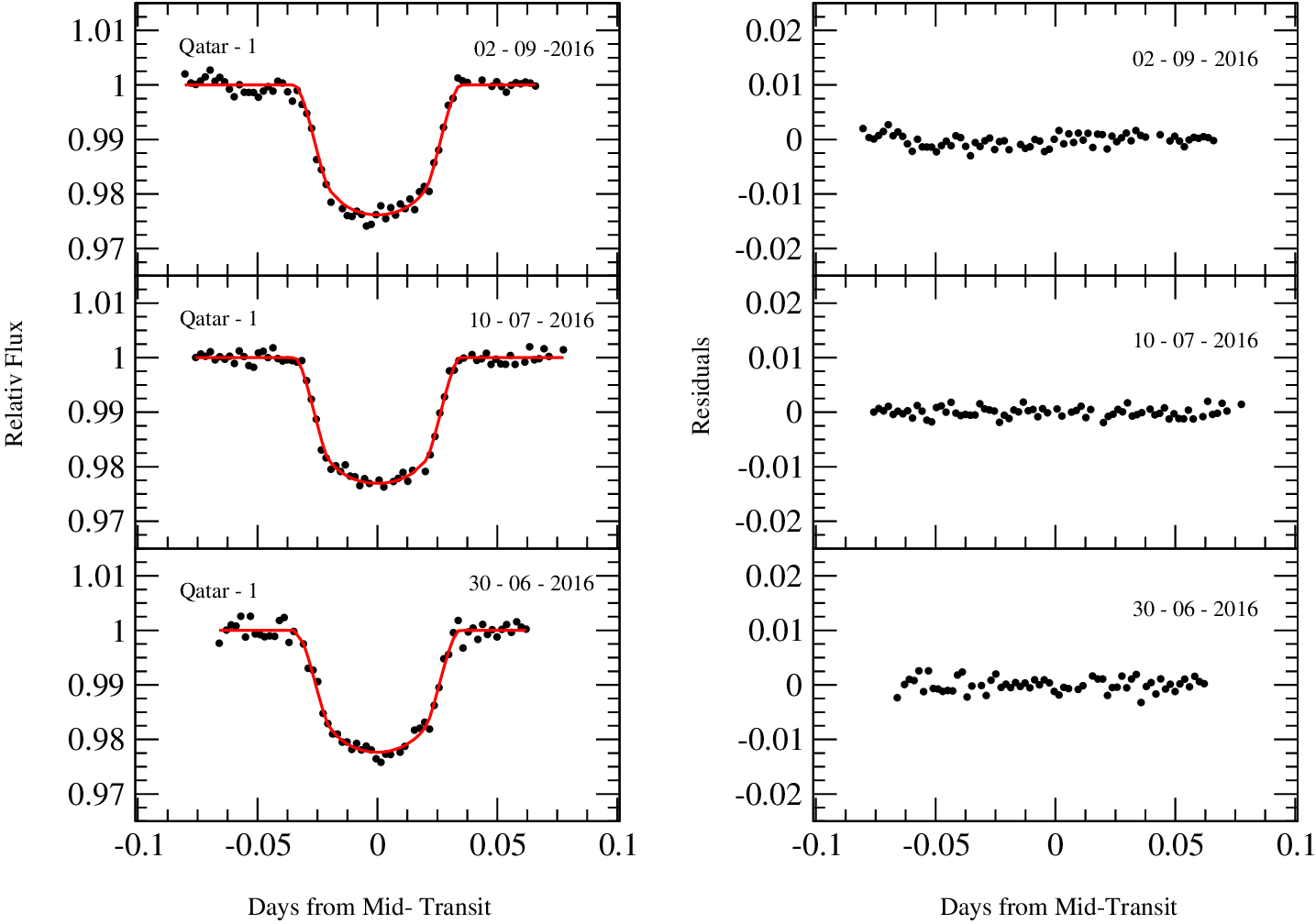}
\caption{Left panels: the normalized relative flux as a function of the time (the offset from the mid-transit time in the Barycentric Julian
Date (BJD) with the time standard Barycentric Dynamical Time (TDB), i.e., TDB-based BJD) of three new light curves used in this work: points are the data and curves are the models. Right panels: the corresponding residuals.}
\end{center}
\end{figure}
\begin{table}[h]
\caption{ Results of best fitted parameters using TAP. \label{table1}}
\small
\begin{center} 
\begin{tabular}{llll}
\toprule
\toprule
{Parameters} & {30-06-2016} & {10-07-2016} & {02-09-2016} \\
\midrule
Orbital Inclination (i$^\circ$) & 84.23\textsupsub{+0.059}{  -0.059} & 84.30\textsupsub{+0.120}{  -0.120} & 84.28\textsupsub{+0.130}{  -0.130} \\  
\\[-1em]
Scaled Semi-major axis (a/R\textsupsub{}{$\ast$}) & 6.335\textsupsub{+0.059}{ -0.059} & 6.348\textsupsub{+0.053}{ -0.053} & 6.323\textsupsub{+0.055}{ -0.055} \\  
\\[-1em]
Planet-to-Star radius ratio (R\textsupsub{}{p}/R\textsupsub{}{$\ast$}) & 0.1431\textsupsub{+0.0028}{ -0.0029} & 0.1461\textsupsub{+0.00022}{ -0.00021} & 0.1477\textsupsub{+0.0021}{ -0.0022} \\ 
\\[-1em]
Mid-transit time (T\textsupsub{}{m}) & 570.34628\textsupsub{+0.00050}{\  -0.00050} & 580.28552\textsupsub{+0.00022}{\  -0.00021} & 634.24666\textsupsub{+0.00035}{\  -0.00034} \\ 
\\[-1em]
Linear limb-darkening coefficient (u\textsupsub{}{1}) & 0.567\textsupsub{+0.047}{ -0.047} & 0.512\textsupsub{+0.045}{ -0.045} & 0.520\textsupsub{+0.046}{ -0.047} \\  
\\[-1em]
Quadratic limb-darkening coefficient (u\textsupsub{}{2}) & 0.166\textsupsub{+0.049}{ -0.048} & 0.130\textsupsub{+0.048}{ -0.48} & 0.195\textsupsub{+0.048}{ -0.048} \\  
\\[-1em]
Sigma red & 0.0029\textsupsub{+0.0031}{ -0.0020} & 0.001\textsupsub{+0.0013}{ -0.00071} & 0.0044\textsupsub{+0.0016}{ -0.0016} \\ 
\\[-1em]
Sigma white & 0.0012\textsupsub{+0.00023}{ -0.00028} & 0.00068\textsupsub{+0.000076}{ -0.000074} & 0.00057\textsupsub{+0.00014}{ -0.00020} \\ 
\\[-1em]
\bottomrule
\end{tabular} 
\end{center} 
\end{table} 
\section{Transit Timing Variation Analysis using O-C Diagram}
To investigate possible TTVs, it necessary to make an O-C diagram. This shows the difference between the observed mid-transit time, T\textsupsub{}{m}, and the calculated mid-transit time T\textsupsub{c}{m}. In this regard, we first calculated T\textsupsub{c}{m} for each epoch (E) using the best fitted values of P and T\textsupsub{}{0} (see Section 4) and then  plotted the O-C diagram as a function of epoch in Fig. 3. We found that the null-TTV model provides better fit to the O-C data with $\chi$\textsupsub{2}{red}= 1.12. This suggest that there is no evidence of TTVs, from which we conclude that there is no evidence of additional planets in the Qatar-1 system either. Our result agrees with those reported in Maciejewski et al. (2015) and Collins et al. (2017).\\ \\
\begin{figure}[h]
\begin{minipage}{8cm}
\centering
\includegraphics[width=8cm]{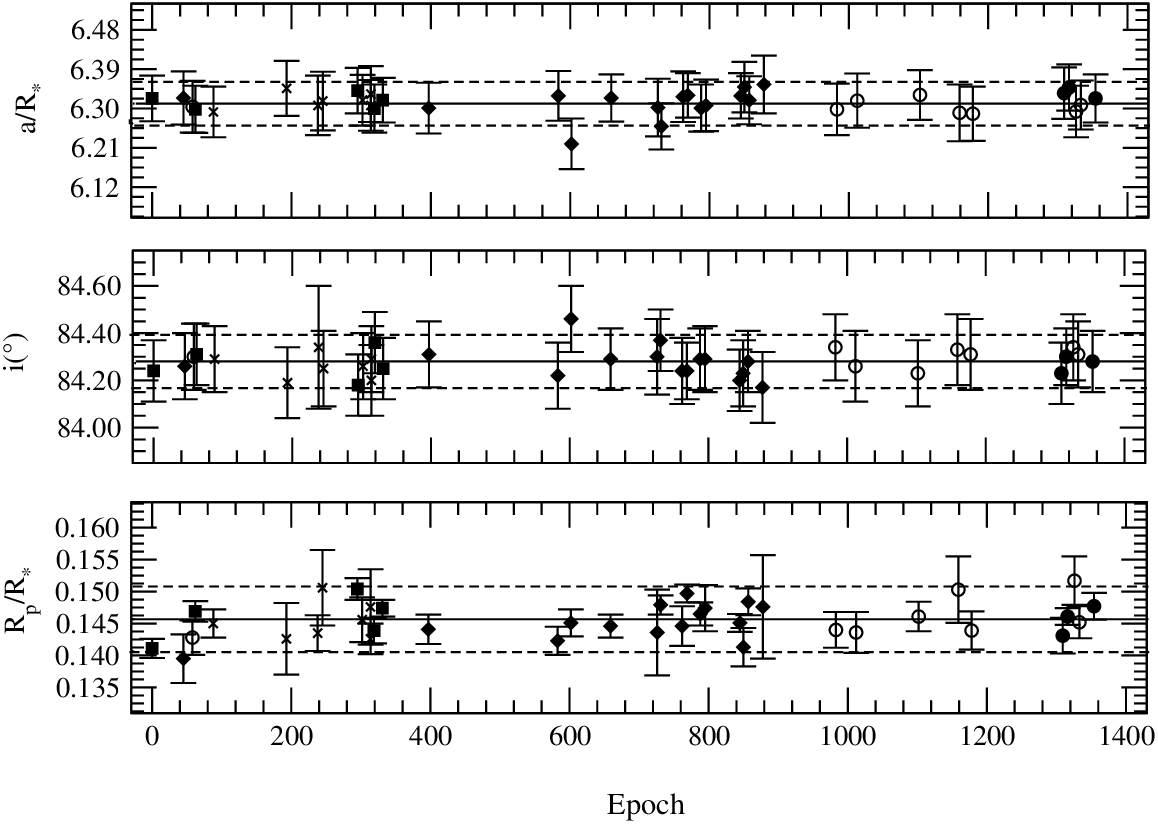}
\caption{Distribution of i, a/R\textsupsub{}{$\ast$} and R\textsupsub{}{p}/R\textsupsub{}{$\ast$} as a function of epoch. In this figure squares are used for data from Covino et al. (2013), diamonds for Maciejewski et al. (2015), crosses for Von Essen et al. (2013), open circles for ETD and filled circles for the data observed by us. The continuous lines denote weighted mean values. Dashed lines denote the uncertainties at the confidence level of 95.5$\%$ (i.e. 2$\sigma$), where $\sigma$ is the weighted standard deviation. \scriptsize \label{fig_2}}
\end{minipage}
\hfill
\begin{minipage}{8cm}
\centering
\includegraphics[width=8cm]{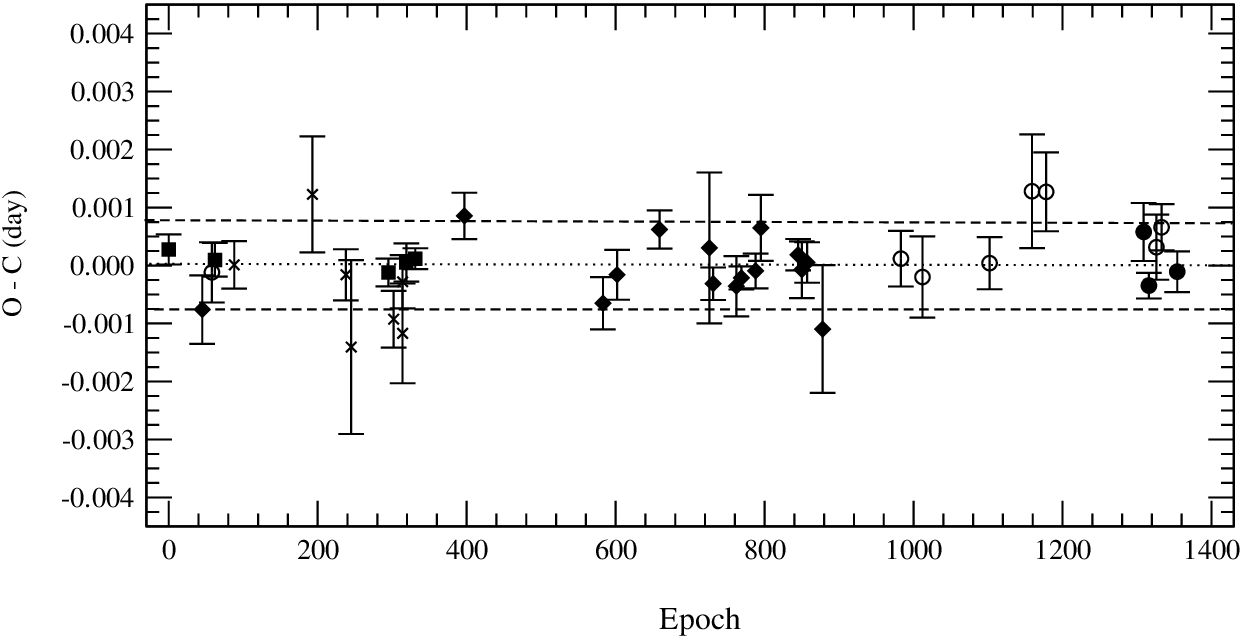}
\caption{Residuals of the transit times from the refined linear ephemeris. Square  symbols are used for the data from Covino et al. (2013), diamonds for Maciejewski et al. (2015), crosses for Von Essen et al. (2013), open circles for ETD and filled circles for the new transits data of this work. The dotted lines denote null TTV model. Dashed lines denote the uncertainties at the confidence level of 95.5$\%$ (i.e. 2$\sigma$), where $\sigma$ is the weighted standard deviation. \scriptsize \label{fig_3}}
\end{minipage}
\end{figure}
\section{Concluding Remarks}
The three new transit light curves of the Qatar-1 system are analyzed. Together with these three light curves, those available in literature are all further analyzed with the same procedure for a uniform estimation of the physical and orbital parameters. All the determined parameter values are consistent with previous works.  Our result suggests that there is no evidence of TTVs in the Qatar-1 system, which allows us to conclude that there may not be an additional planet present in this system. However, this result should be further confirmed by high-cadence, high-precision follow-up observations using the 3.6-m DOT (ARIES, Nainital, India.)
\section*{Acknowledgments}
{The authors thank the anonymous referee for the suggestion to improve this manuscript. We also thank to UGC, New Delhi for providing the financial support through Major Research no. UGC-MRP 43-521/2014(SR). Observation times granted by the HCT time allocation committee is gratefully acknowledged. PT expresses his sincere thanks to IUCCA, Pune for providing the supports through IUCCA Associateship Programme. The facilities provided by the Dept. of Pure \&\ Applied Physics, Guru Ghasidas Central University, Bilaspur (C.G.), India, is also gratefully acknowledged.}
%
%
\beginrefer
\refer Agol E., Steffen J., Sari R., Clarkson W. 2005, MNRAS, 359, 567

\refer Collins K. A., Kielkopf J. F., Stassun K. G. 2017, AJ, 153, 78

\refer Covino E., Esposito M., Barbieri M. et al. 2013, A\&A, 554, A28

\refer Gazak J. Z., Johnson J. A., Tonry J. et al. 2012, AdAst, 697, 967

\refer Holman M. J., Murray N. W. 2005, Science, 307, 1288

\refer Holman M. J., Fabrycky D. C., Ragozzine D. et al. 2010, Science, 330, 51

\refer Maciejewski G., Fernandez M., Aceituno F. J. et al. 2015, A\&A, 577, A109

\refer Mandel K., Agol E. 2002, ApJ, 580, L171

\refer Von Essen C., Schroter S., Agol E., Schmitt J. H. M. M. 2013, A\&A, 555, A92

\endrefer           

\end{document}